\documentclass[%
 reprint,
 amsmath,amssymb,
 aps,
floatfix,
]{revtex4-1}

\usepackage{natbib}
\newcommand{\mean}[1]{\ensuremath{\left\langle #1 \right\rangle}}  
\usepackage{graphicx}
\usepackage{dcolumn}
\usepackage{bm}

\begin{document}


\title{An \textit{In Vitro} Nematic Model for Proliferating Cell Cultures}


\author{Sunil Pai}
\author{Nathan Loewke}
\author{Morgaine Green}
\author{Christine Cordeiro}
\author{Elise Cabral}
\author{Bertha Chen}
\author{Thomas Baer}
\affiliation{Stanford University}


\date{\today}

\begin{abstract}
Confluent populations of elongated cells give rise to ordered patterns seen in nematic phase liquid crystals. We correlate cell elongation and intercellular distance with intercellular alignment using an amorphous spin glass model. We compare \textit{in vitro} time-lapse imaging with Monte Carlo simulation results by framing a novel hard ellipses model in terms of Boltzmann statistics. Furthermore, we find a statistically distinct alignment energy at quasi-steady state among fibroblasts, smooth muscle cells, and pluripotent cell populations when cultured \textit{in vitro}. These findings have important implications in both non-invasive clinical screening of the stem cell differentiation process and in relating shape parameters to coupling in active crystal systems such as nematic cell monolayers.
\end{abstract}


\maketitle


\section{Introduction}

Self-organization is an emerging field in biophysics with important implications in tissue engineering, biomechanics, and regenerative medicine. Embryo growth and muscle structure result from such self-organization processes. We develop biophysical models to describe this non-equilibrium process that leads to an ordered biomaterial. Our approach provides deeper understanding of how tissue structure arises from the individual cells as they proliferate, change morphology, and increase in density.

Histology samples and time-lapse imaging indicate that certain cells (\textit{nematoid} cells) tend to align their major axes with those of neighboring cells. The direction of the orientation vector for each cell (which lies along the major axis of an ellipse fit to the cell contour) is usually determined by tensile mechanical force dipole interactions along cell cytoskeletons \cite{gjorevski2015dynamic,ramaswamy2010activematter}. The interactions among the actin fibers and cadherin anchoring junctions of the cells are responsible for the cell intercalation, adhesion, and axially-aligned migration that describe the motion and behavior of a large variety of cell types including fibroblasts, smooth muscle cells (SMCs), and osteoblasts \cite{gruler1999nematic,gjorevski2015dynamic}. The biomechanical organization of cells in a petri dish resembles 2D liquid crystals at a quasi-steady state, and has been previously studied using Langevin dynamics model \cite{kemkemer2000elastic} and pairwise elastic interaction theories applied to microtubule ensembles \cite{decamp2015orientational}, mouse fibroblast monolayers \cite{duclos2014perfect}, and bioengineered nematic vesicles \cite{keber2014topology}.

We extend previous theoretical and experiment-based studies employing elastic continuum theory \cite{gruler1999nematic,kemkemer2000elastic} and active nematics spin glass \cite{bischofs2005effect,duclos2014perfect,decamp2015orientational,keber2014topology} to understand how order emerges in such non-equilibrium systems. Our model predicts, and our time-lapse experiments confirm, vortex structures in cell monolayers which evolve and interact amongst each other during the phase transition from a disordered to ordered state, typical of active nematics \cite{menzel2014active,ramaswamy2010activematter,shi2013topological}.

To extend existing active nematics studies, we develop a nematic cell alignment theory (NCAT) which relates cell elongation to cell alignment at maximal cell packing density, at which point the active nematic reaches a quasi-steady state. Incorporating the analogy between nematoid cell arrangements \textit{in vitro} and 2D nematic crystals, we model the cells as mobile, interacting ellipses with coupling interactions that increase as they proliferate to higher densities. Specifically, this increased probability of interaction among cells leads to higher alignment correlation. This fundamental self-organization phenomenon converges to statistically distinct final energies at high density for different cell phenotypes due to variations in cytoskeletal interactions that affect cell elongation. In this paper, we show how NCAT combines statistical mechanics and a hard ellipse model in a new way to relate cell shape to cell-to-cell interactions and ultimately explain \textit{in vitro} cell monolayer self-organization.

\section{Theory} \label{morphotypes}
\subsection{Nematic Cell Alignment Theory}
We propose a nematic cell alignment theory (NCAT) that describes the evolution of cell alignment via pairwise cell-cell interactions, which vary with cell density and cell elongation.

We define cells as ellipses with orientations in the range $ \theta_i \in [0,\pi) $ as the unit vector orientation of the major axis of the ellipse. The anisotropic interaction term between two cells $ i $ and $ j $ is  proportional to $ \cos(2(\theta_i- \theta_j)) $ and occurs primarily for neighboring cells \cite{schwarz2002elastic}. To simplify our model, we assume that cell migration is slow enough such that the cell interactions are minimally affected by their migration trajectories, an assumption made in previous literature for spin lattice studies \cite{toner1998flocks}. Finally, the model does not include any other repulsive or attractive potentials among the cells, instead assuming that cell-cell interactions are sufficiently absorbed in the NCAT Hamiltonian.

Our simplified NCAT Hamiltonian is the Lebwohl-Lasher model \cite{lebwohl1972nematic}:
\begin{align} \label{eq:hamiltonian}
\mathcal{H} &= -A \sum_{\mean{i,j}} \cos 2(\theta(\mathbf{r}_i) - \theta(\mathbf{r}_j))
\end{align}
where $ A $ is the coupling term influenced by cell alignment interaction strength (which correlates with cell eccentricity) and cell density. $ \mathbf{r}_i $ represents the position of cell $ i $. $ \mean{i,j} $ is the indicator for a valid cell neighbor pair. $ \theta(\mathbf{r}) $ is the cell's major axis orientation at position $ \mathbf{r} $. 

This Hamiltonian employs the same order parameter as that used in the literature \cite{gruler1999nematic}, $ \cos 2 \theta $, which accounts for the degeneracy between $ \theta $ and $ \theta + \pi $ orientation. A key difference is that Equation \ref{eq:hamiltonian} employs a pairwise (bond) energy rather than a per-cell energy very much in the spirit of \cite{friedrich2011nematic}, and this avoids the pitfall of defining an ad hoc director $ \mathbf{n} $ for the cell monolayer (i.e. a set of predefined angles for all cells to align to).

The partition function for the cell orientation is defined as (letting $ 1/kT = 1 $):
\begin{align}
Z &= \sum_{\Theta} \exp\left(A \sum_{\mean{i,j}} \cos 2(\theta(\mathbf{r}_i) - \theta(\mathbf{r}_j))\right)
\end{align}
We define the cell locations as $ \{\mathbf{r}_1,\mathbf{r}_2,\ldots,\mathbf{r}_N\} $ and $ \Theta $ as the set of valid microstates comprised of the cell orientations at the lattice points $ \theta(\mathbf{r}_i) $.

In NCAT, we model cells as a network ensemble of \textit{interacting hard ellipses} with eccentricity $ \epsilon $ and cell density $ \rho $, both measurable experimental parameters that determine the magnitude of $ A $. We assume that all neighboring cells may contact each other at any point along their perimeters with equal probability. We calculate the energy $ \hat u(\epsilon) $ for two interacting ellipses by averaging $\cos 2(\theta - \theta')$ over all possible tangent cell contact points (see Figure \ref{fig:disclination_exp}(c)). Next, we calculate the same bond energy $ u(A) $ for two cells using $ \mathcal{H} $ from Eq. \ref{eq:hamiltonian} (ignoring the summation). The resulting expressions for $ \hat u $ and $ u $ (derived in Appendix C of Supplemental Material) are:
\begin{align}
\hat u(\epsilon) = \left(\frac{\sqrt{1 - \epsilon^2} - 1}{\sqrt{1 - \epsilon^2} + 1}\right)^2 & & u(A) = \frac{I_1(A)}{I_0(A)}
\end{align}
where $ I_n $ is the modified Bessel function of the first kind. We numerically evaluate $ A(\epsilon) = u^{-1}(\hat u (\epsilon)) $ to determine the coupling parameter $ A $ in terms of the average eccentricity of the cells in the time-lapse experiments for comparison with simulation results.

\subsection{Simulation Design and Results}

\begin{figure}
\includegraphics[width=0.45\textwidth]{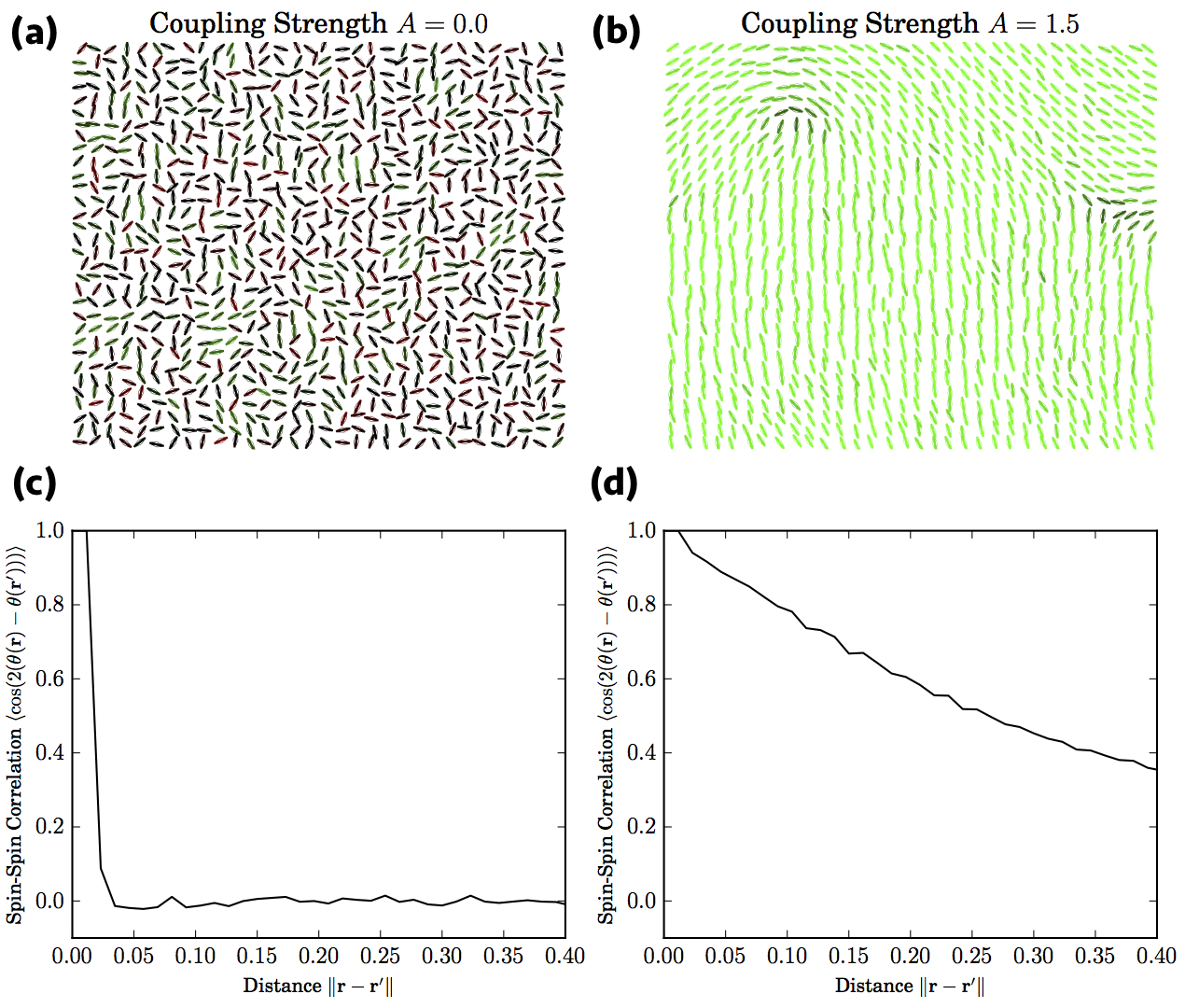}
\includegraphics[width=0.45\textwidth]{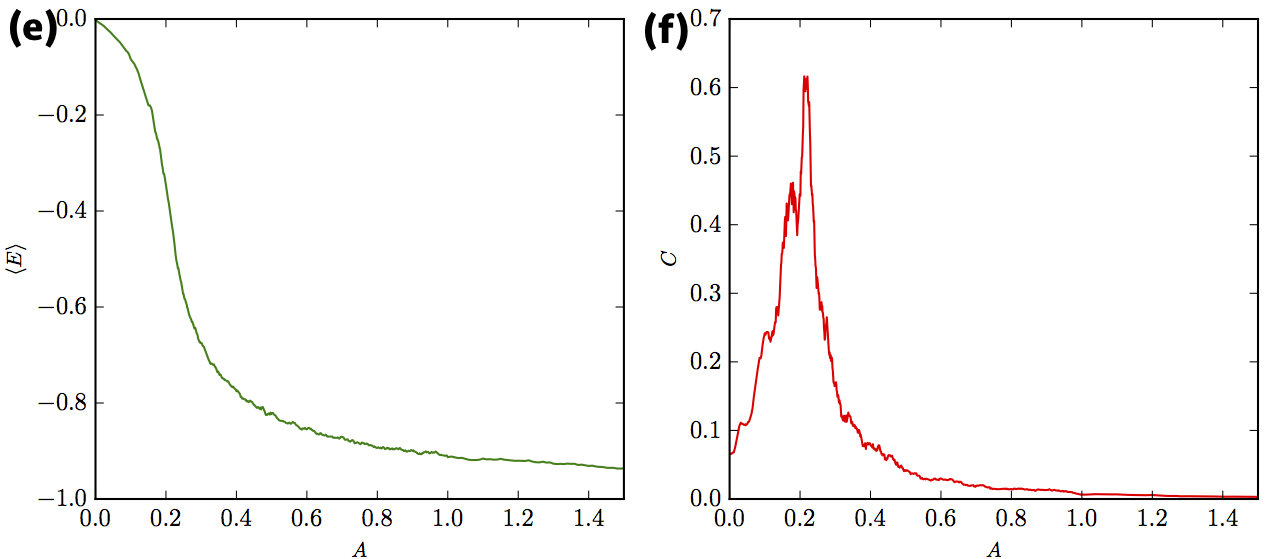}
\caption{\label{simul_demo} Monte Carlo simulation for NCAT model at nematic, quasi-steady state regime showing lattice orientations and correlation vs distance plots at isotropic, low $ A $ (a,c) and at nematic, high $ A $ (b,d). Note the $ m = \pm 1/2 $ disclinations in (b). Continuous phase transition occurs when varying $ A $, which can be seen in (e) the average bond energy and (f) the time-averaged heat capacity ($ C \propto A^2 (\mean{E^2} - \mean{E}^2) $) at quasi-steady state. Both plots are smoothed using low pass filter and correspond to simulations with a 5-nearest-neighbor connection rule with fixed lattice points.}
\end{figure}

Using Monte Carlo simulation, we model critical behavior allowing cells to transition from an unordered, isotropic phase to an ordered, nematic phase as they proliferate and increase their contact area. We assume that the cells are self-propelled particles that reach a linearly stable configuration as predicted by \cite{menzel2014active}. The primary purpose of simulation is not necessarily to capture dynamics, but rather to run the simulation over a range of $ A $ to quasi-steady state (at confluent density) and plot the average energy per lattice bond while varying the cell eccentricity. Our simulation uses 1024 lattice sites with periodic boundary conditions and starts with randomly aligned cells (Figure \ref{simul_demo}(a)) and at quasi-steady state reproduces localized vortices typical for apolar nematic media, (Figure \ref{simul_demo}(b)).

The phase transition in Figure \ref{simul_demo} (e,f) is similar in nature to the Kosterlitz-Thouless (KT) transition observed for the $ XY $ lattice model \cite{kemkemer2000elastic,chate2006simple}. This transition is manifested by the change in the spin-spin correlation dependence which shifts from exponential to power law \cite{kosterlitz1974critical,chate2006simple} in Figure \ref{simul_demo}(c,d) and at the peak of the time-averaged heat capacity in Figure \ref{simul_demo}(f). This approach to long range order may be observed both experimentally from cell culture experiments and theoretically from simulation. At the critical $ A $, vortices stop being generated and the gyrating structures typically observed in nematic cell cultures begin to be observed.

\begin{figure}
\begin{center}
\includegraphics[width=0.23\textwidth]{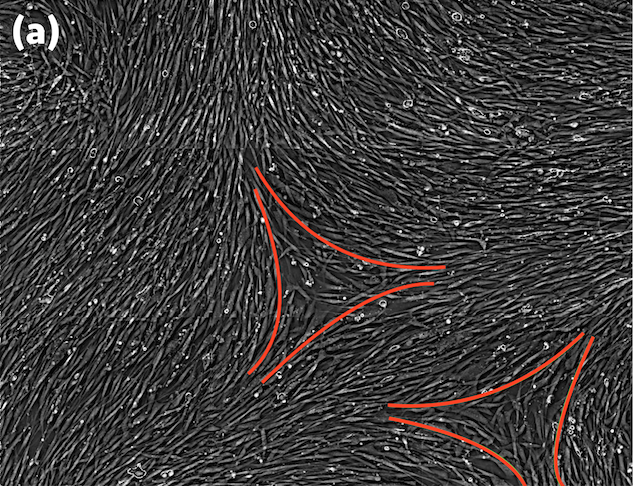}
\includegraphics[width=0.215\textwidth]{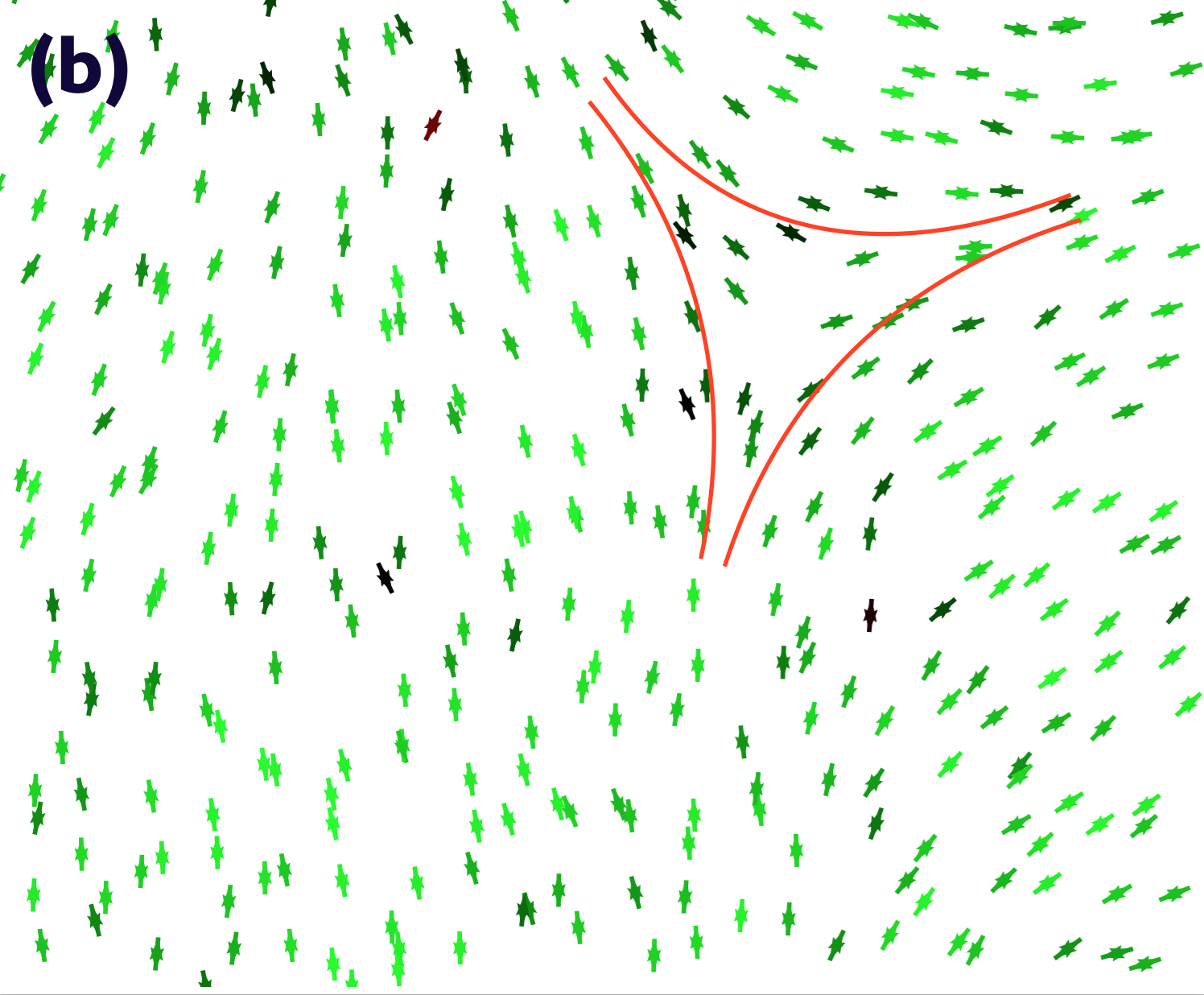}\\
\includegraphics[width=0.48\textwidth]{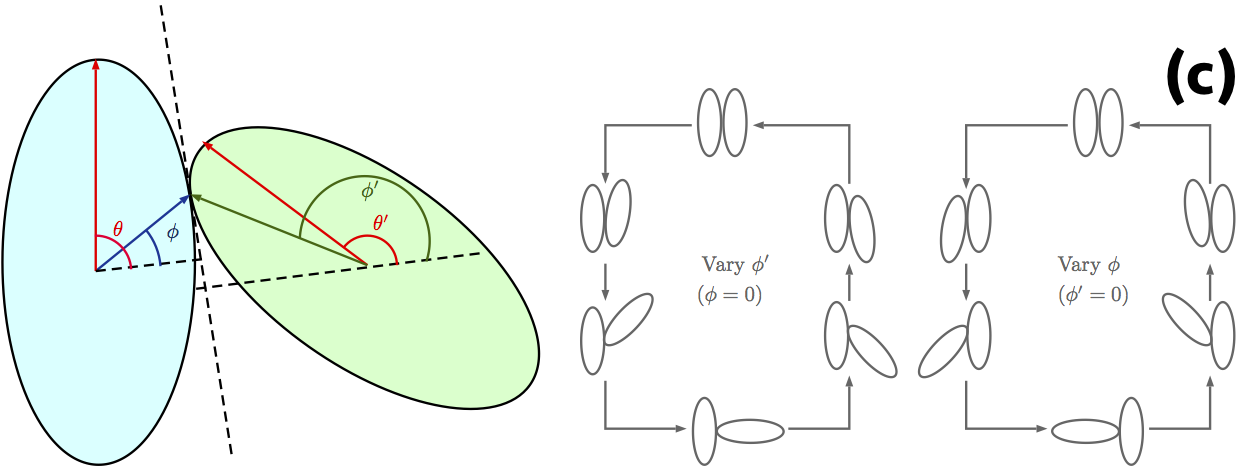}
\end{center}
\caption{The presence of $ m = 1/2 $  disclinations in experiment (fibroblast/SMC mixture) (a) and in simulation (b). (c) Hard ellipse interaction as it relates to alignment differences between the major axes. Note the role eccentricity plays in the alignment. See appendix C of Supplemental Material for derivations.}
\label{fig:disclination_exp}
\end{figure}

To expose the $ m = \pm 1/2 $ vortex structures typically observed in elastic continuum theory (see Appendix B of Supplemental Material) \cite{kemkemer2000elastic}, we include a velocity term which allows the lattice sites to move. These vortex patterns observed experimentally and in simulation (see Figure \ref{fig:disclination_exp}) demonstrate that while cell monolayers typically do not reach equilibrium, the macrostate energy with the vortices are close to non-equilibrium steady state. Note that $ m = 1/2 $ vortex usually interacts with a $ m = -1/2 $ vortex in both simulation (when lattice sites are allowed to move along their orientations) and experiment. This interaction can be seen in simulation in Figure \ref{simul_demo}(a) and \textit{in vitro} in Figure \ref{fig:energies_comp}(b).

\section{Materials and Methods}
In the \textit{in vitro} experiments, cells were plated at approximately equal density and under \textit{highly controlled incubation conditions} on an incubated 12-well culture system (see Appendix A of Supplemental Material). We studied mixtures of iPSC (induced pluripotent stem cells), smooth muscle cells (SMC), and Huf3 fibroblast (FB) cell populations. These are important cell populations involved in differentiation of stem cells to smooth muscle cells. The purpose of varying the mixtures is to correlate the NCAT model parameters with mixtures of different cell phenotypes that may appear during differentiation. The purpose of this experiment was to show that there was a significant difference in the long range order of cell alignment that may allow one to better understand biomechanics involved in such a differentiation process.

To image the cells, we developed an automated quantitative phase time-lapse microscope to quantify the dry mass profile in the image (which increases the index of refraction relative to water) \cite{popescu2006diffraction}. The images were taken every 10 minutes over the course of 20-30 hours. We then processed the images to measure the eccentricity and orientation of each cell over the course of the experiment.

We find the cell locations $\mathbf{r_i}$ using local regional maxima. Average eccentricity data was computed by using automatic cell segmentation and contour analysis techniques. Finally, we employed \textit{histogram of oriented gradients} (HOG) \cite{dalal2005histograms} to compute the cell orientation direction at each cell location. All image post-processing was performed using the standard \texttt{skimage} implementation in Python. All Monte Carlo simulations were run in Python on 16-core computers to parallelize the simulation for different values of $ A $.

\section{Results and Discussion}

\begin{figure}
\centering
\includegraphics[width=0.22\textwidth]{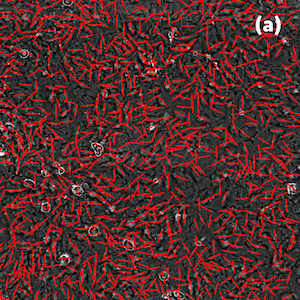}
\includegraphics[width=0.22\textwidth]{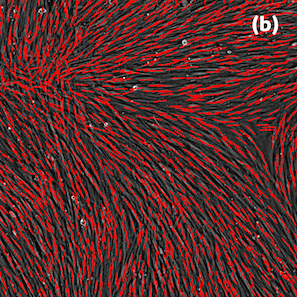}\\
\includegraphics[width=0.23\textwidth]{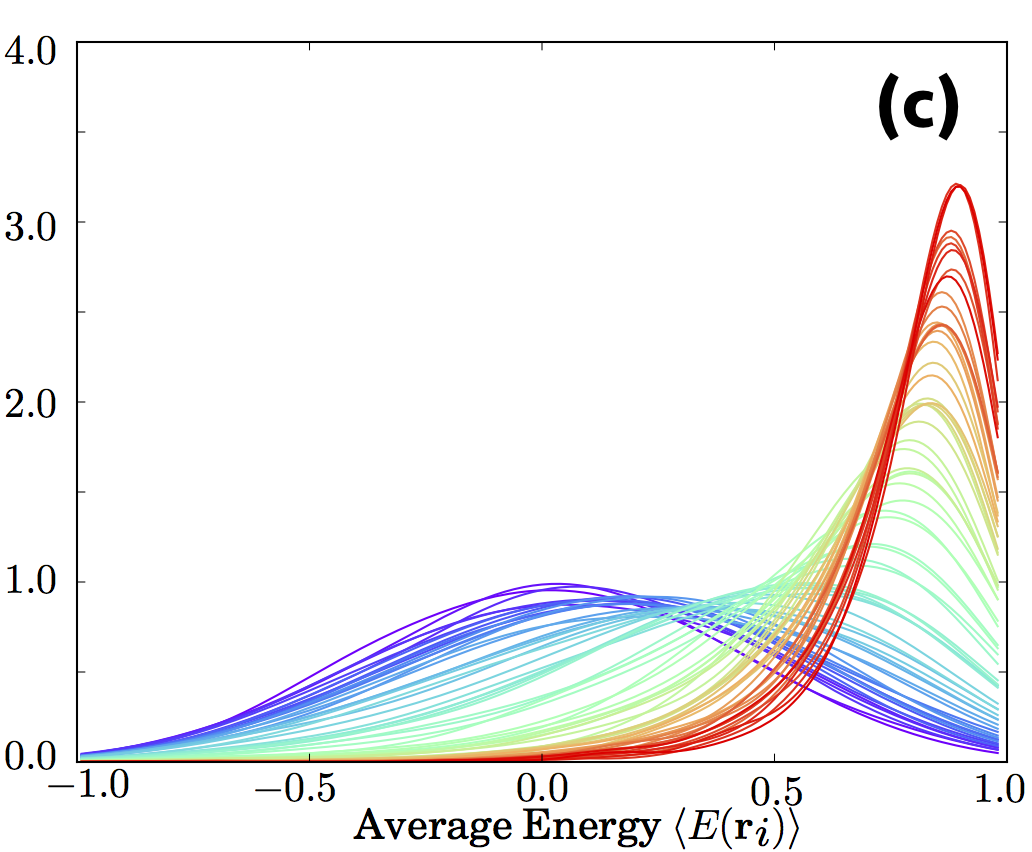}
\includegraphics[width=0.23\textwidth]{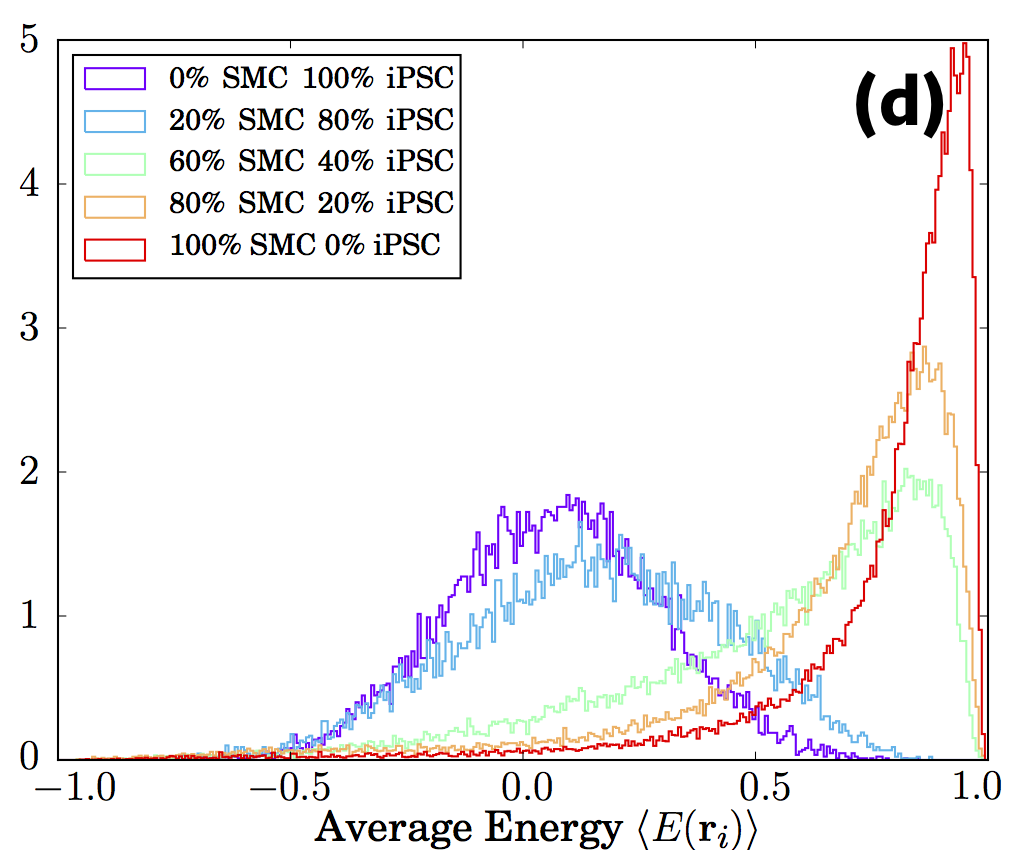}
\caption{\label{angular_dist} There are differences in alignment between dermal iPSC (a) and FB (b), shown using time-lapse images overlaid with cell orientation directions. We also analyze average cell energy (average correlation of cells with neighbors) using simulation in the range $ A = 0.05 $ (violet) to $ A = 0.7 $ (red) in (c). An experimental histogram of average energy for mixtures of SMC/iPSC cells in quasi-steady state in (d) shows similar distribution to simulation.}
\label{fig:energies_comp}
\end{figure}

We find distinguishable alignment patterns in timelapse images that result from variations in cell eccentricity for different phenotypes. Figure \ref{fig:energies_comp} shows the qualitative differences in our analysis of fibroblasts and pluripotent cells.

To verify that the distribution of alignment energies in our \textit{in vitro} experiments can be compared to the NCAT simulation, we first analyze the energy histogram on a per-cell basis. A comparison of simulated and experimental histograms for SMC/iPSC cell combinations are shown in Figure \ref{fig:energies_comp} (c) and (d) respectively.


The NCAT analysis allows us to distinguish differences in mixtures of binary cell populations and agrees with our simulation results.

\begin{figure}
\centering
\includegraphics[width=0.24\textwidth]{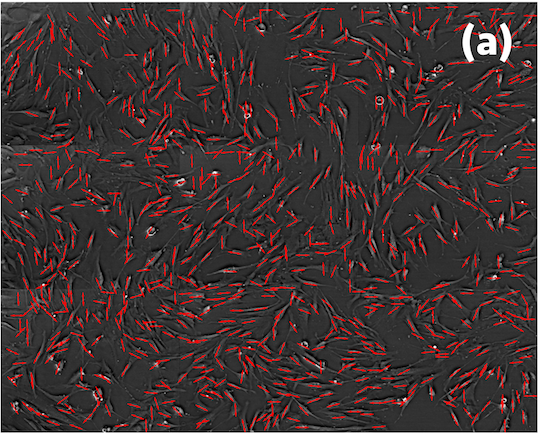}%
\includegraphics[width=0.24\textwidth]{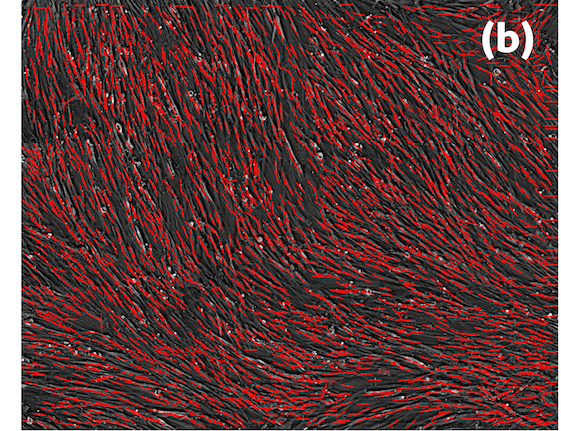}\\
\includegraphics[width=0.24\textwidth]{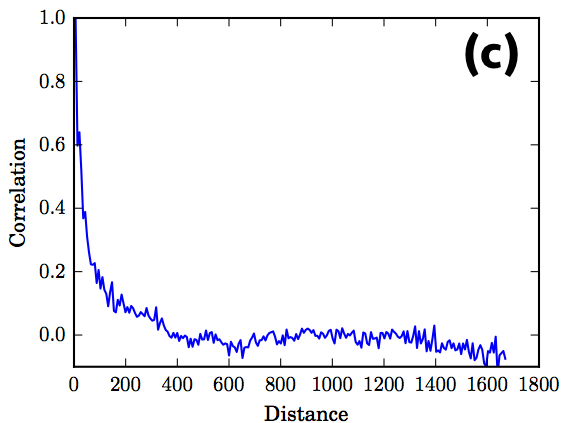}%
\includegraphics[width=0.24\textwidth]{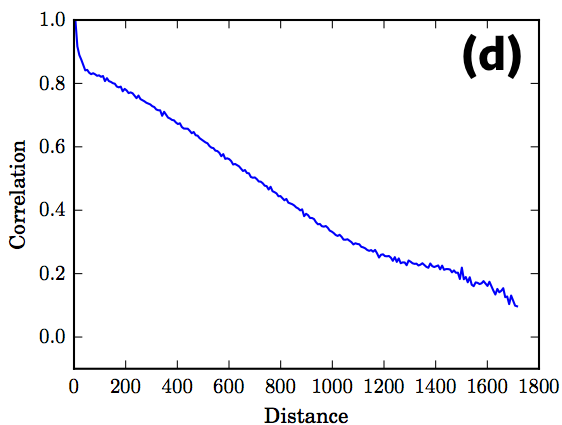}%
\caption{\label{fig:analyzed_smc} Nematic ordering in a proliferating smooth muscle cell-fibroblast mixture from (a) unordered, low-density to (b) ordered, high density with corresponding spin-spin correlation function behavior in (c) and (d) respectively. Note the evolution from exponential to power law behavior (in this case, linear due to presence of vortices).}
\end{figure}

Figure \ref{fig:analyzed_smc} (c) and (d) shows the experimental transition from uncorrelated to correlated cell orientations for nematoid cells, which compares well with our simulation results in Figure \ref{simul_demo} (c) and (d).\footnote{The Supplemental Material contains time-lapse videos of FB, SMC, and iPSC using the procedure in Materials and Methods. Also plotted is the total angular correlation of the cells in the samples as a function of distances calculated for each video frame. As the FB and SMC cells become more confluent, the angular correlation function changes from an exponential to a linear dependence on distance, however the iPSC cells even at high densities maintain their original exponential dependence with cell-to-cell distance.}

Using energy and correlation data from experiment and simulation, we can approximate the critical eccentricity and density at which nematic ordering occurs for nematoid cell mixtures. 

\begin{figure}
\centering
\includegraphics[width=0.48\textwidth]{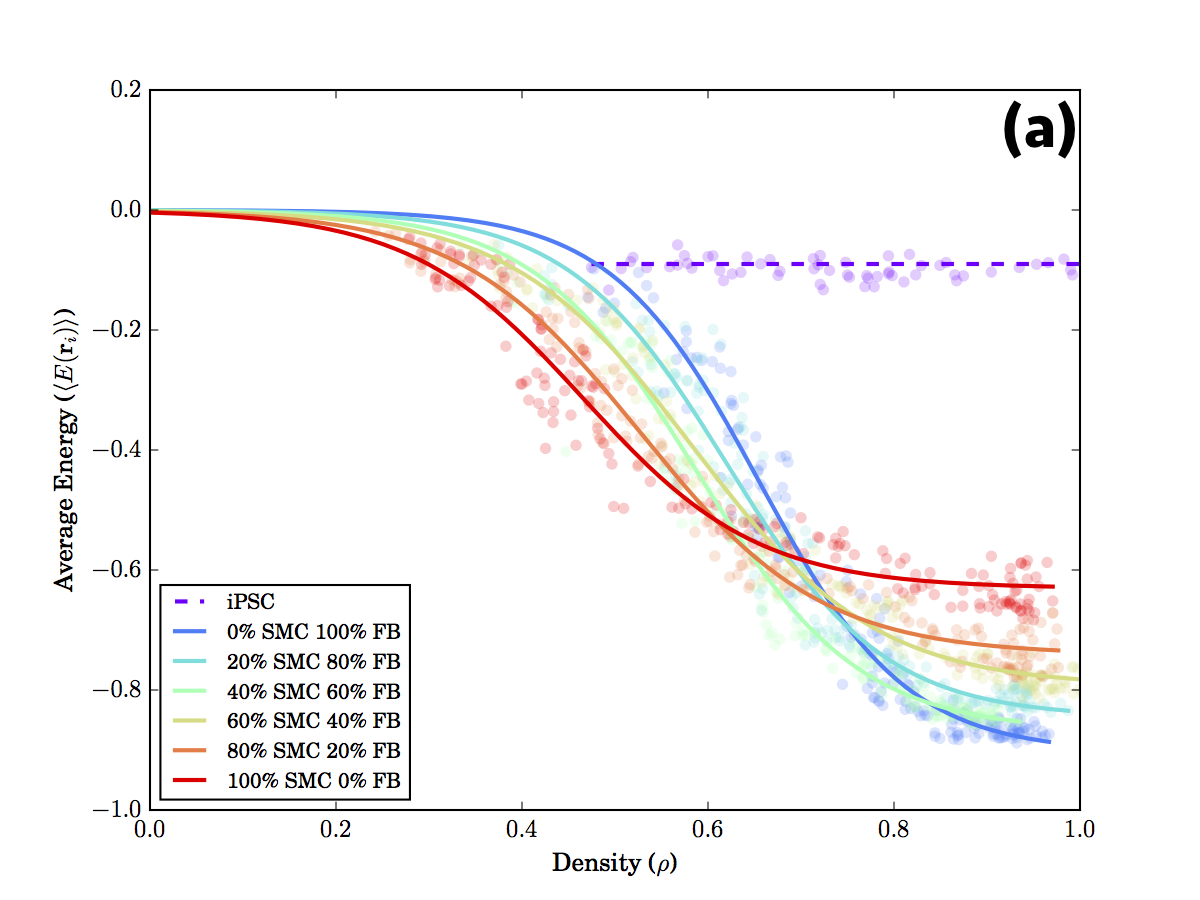}\\
\includegraphics[width=0.48\textwidth]{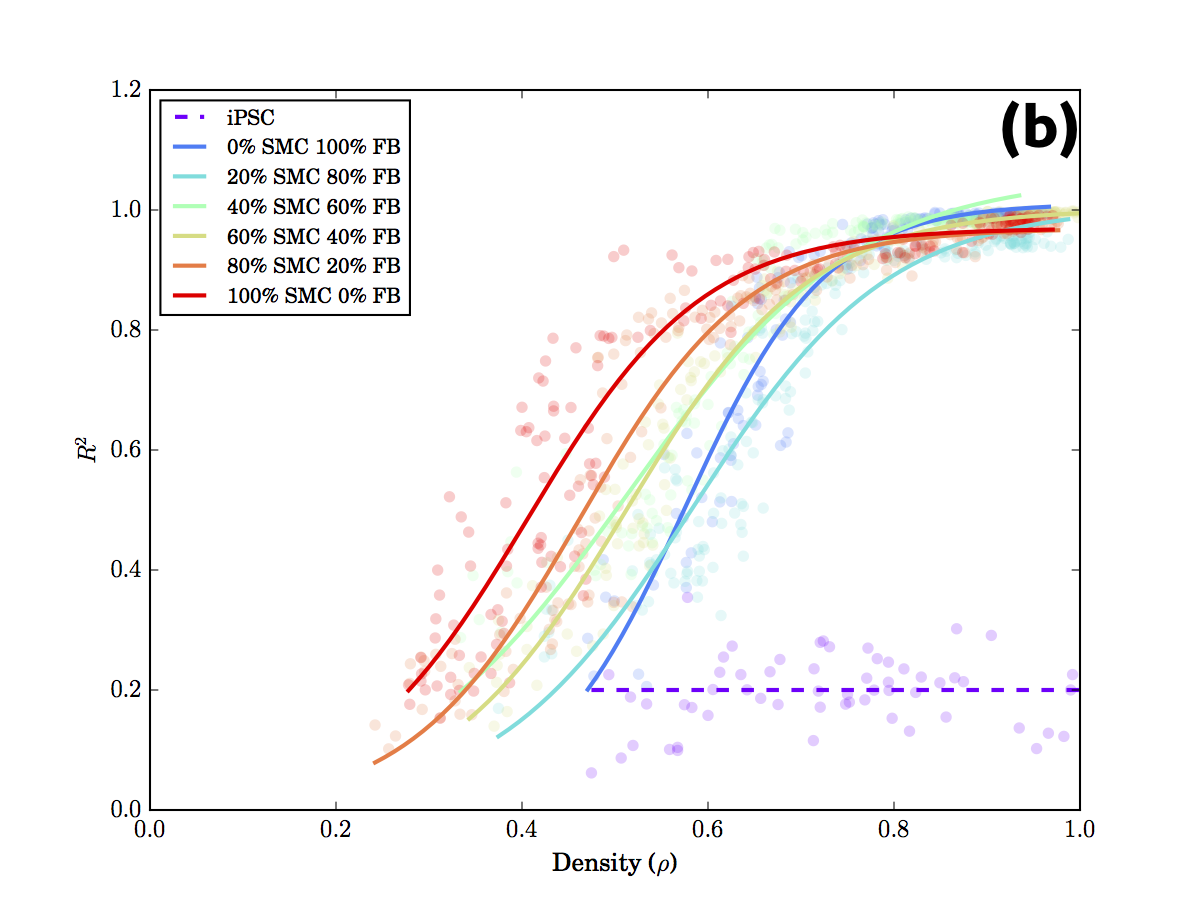}\\
\includegraphics[width=0.48\textwidth]{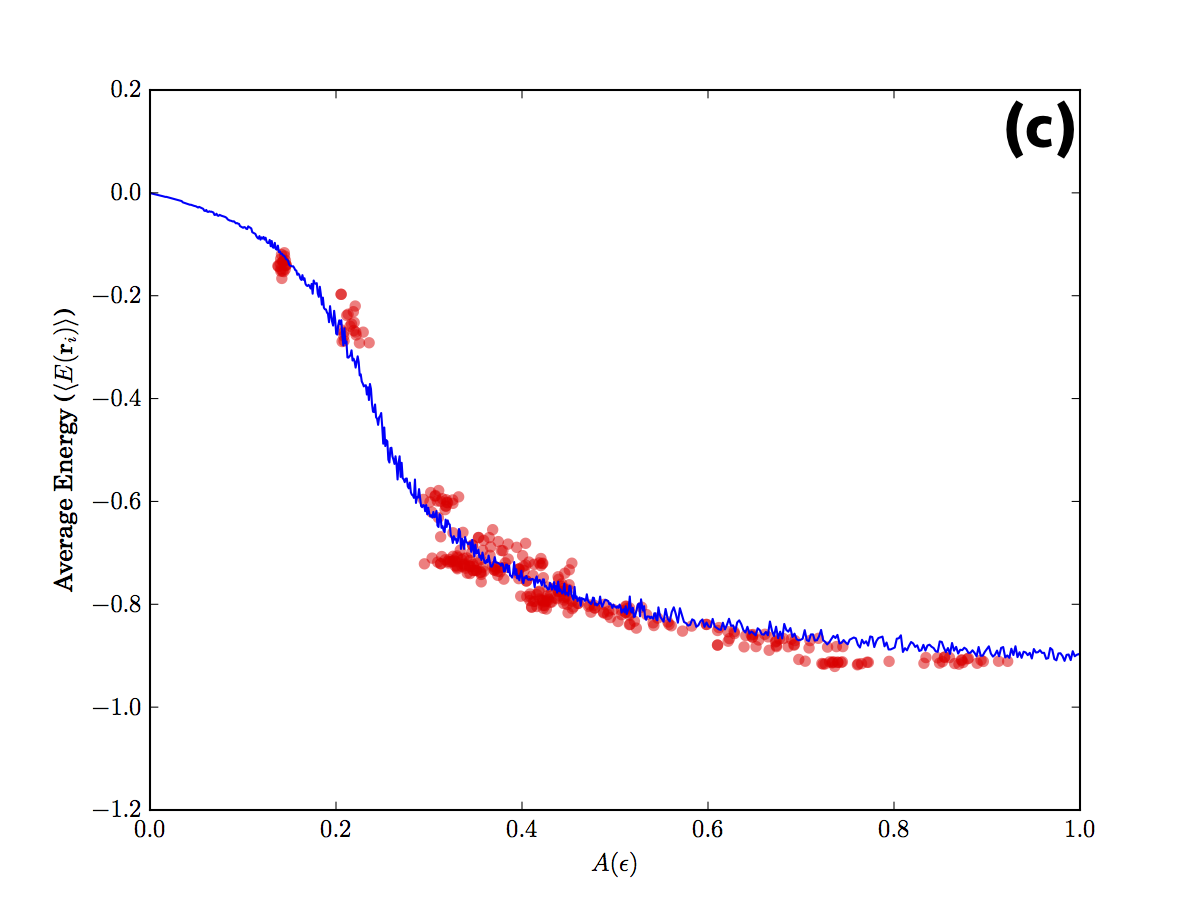}
\caption{(a) Comparison of energies as a function of cell density $ \rho $ for FB/SMC mixtures compared to iPSC, where $ \rho = 1 $ is max density. (b) Phase transition from the same experiment shows an approach to linear correlation behavior (quasi-steady state configuration) with Pearson coefficient $ R^2 $. (c) A comparison of energies at confluency (red points) using $ A(\epsilon) $ (where $ \epsilon $ is the average cell eccentricity in the time-lapse image) and energy predictions from simulation (blue curve). Note: For (a) and (b), solid lines are fits to sigmoid curves to demonstrate overall trends, and do not have mathematical significance.}
\label{fig:phasethryvsexp}
\end{figure}

From Figure \ref{fig:phasethryvsexp} (a), we note that that smooth muscle cells undergo a phase transition at lower densities than the fibroblasts, but reach a higher quasi-steady state energy.

To determine the density at which a phase transition occurs, we track the transition from exponential to power law dependence (shown in Figure \ref{fig:phasethryvsexp} (b)) by plotting the square of the Pearson coefficient $ R^2 $ against the correlation data as a function of density.

The dependence of energy versus coupling constant $ A(\epsilon) $ in the confluent density regime is shown in Figure \ref{fig:phasethryvsexp} (c). We plot average energy for the last 30 frames of time-lapse data for all datasets used in this study (SMC/iPSC and 2 independent FB/SMC experiments), shown in red in Figure \ref{fig:phasethryvsexp} (c). We then ran our Monte Carlo simulations to quasi-steady state and computed the average energy for a range of the coupling parameter $A$. This is shown in the blue solid plot in Figure \ref{fig:phasethryvsexp} (c), showing agreement between NCAT simulation and experiment.

\section{Conclusions}

Based on energy distribution and energy dynamics signatures over the course of a multiwell time-lapse imaging experiment, we observe several trends consistent with NCAT theory and  with recent observations of biological active nematic systems \cite{bischofs2005effect,duclos2014perfect,decamp2015orientational,keber2014topology}. First, as nematic cells proliferate from mid to high density, there is a KT-like phase transition predicted by NCAT and nonequilibrium studies of proliferating active crystals \cite{chate2006simple}. Second, cell cultures typically reach nematic, quasi-steady state configurations with the appearance of $ \pm 1/2 $ vortices, which agrees with our simulations. Finally, mixed cell populations have characteristic quasi-steady state configurations at confluent density that fit to simulation predictions of average energy vs $ A(\epsilon) $.

Our findings introduce eccentricity as a key variable that differentiates the nematic ordering process of fibroblasts, smooth muscle cells, and pluripotent stem cells and correlates with the order parameter. In future investigations, we plan to study the biological basis of cell elongation at confluency and its role in determining $ A $ to further explore the observed correlation between eccentricity and nematic ordering.

Measuring the phenotypic concentration of different cell types in stem cell derived transplants is an important quality control method. This work may provide non-invasive quality assurance measures to correlate model parameters with clinical outcomes for pluripotent stem cell-derived smooth muscle cell-based transplant therapies, where it is important to consider the biomechanical integrity, macro-structure, and phenotypic make-up of cultured cells.

\begin{acknowledgments}
The contributions for this work are as follows: SP developed imaging software, image post-processing, NCAT theory formulation with comparison with experiment, and wrote the paper. TB developed imaging software, oversaw the work, and edited the paper. NL developed image processing routines for quantitative phase imaging. CC optimized the optics for imaging platform. MG and EC performed all biological preparations and time-lapse experiments on imaging platform. BC oversaw the biological preparations and edited the paper.
\end{acknowledgments}

\bibliography{main}

\end{document}


\title{Supplemental Material (Appendix)}

\maketitle

\appendix 

\label{App:AppendixA}
\section{Multiwell Experiments}

For multiwell experiments, we used 1:100 diluted Matrigel in DMEM culture medium. The different cell samples are plated at mid-density (about 50\% confluency, 50000 cells per well). Timelapse images were taken over a period of 3 days for 10-minute intervals. The cells were maintained at standard incubation conditions with a supply of 5\% CO${}_2 $ gas mixture at 37$ {}^\circ $C.

\section{Vortices} \label{App:AppendixB}

The amorphous lattice vortices $m = \pm 1/2 $ can be derived as solutions to the Euler Lagrange equation for the local orientation director $ \mathbf{n}(\mathbf{r}) = (\cos\Phi(x,y),\sin\Phi(x,y)) $, where $ \Phi(x,y) $ is the angle for the local direction in the elastic continuum. In cylindrical coordinates, the equation and solution for $ \Phi(x,y) $ are \cite{kemkemer2000elastic}: 

\begin{align}
0 &= \frac{\partial \Phi}{\partial r^2} + \frac{\partial \Phi}{\partial r} + \frac{1}{r^2}\frac{\partial^2\Phi}{\partial\phi^2}\\
\Phi &= m\phi + \Phi_0
\end{align}

We observe the vortices for $m = \pm 1/2 $ in both cell culture and simulations.

\section{Eccentricity and Coupling Parameter $ A $} \label{App:AppendixC}

We now formalize the relationship between cell eccentricity and coupling parameter $ A(\epsilon) $ by matching the mean energy moments of a simplified two-body hard ellipse interaction and the Boltzmann statistics of the NCAT model. In utilizing this two-body interaction, we perform a relatively simple computation, but we approximate the relationship among the hard ellipses in a lattice network as simply the sum of two-body interactions for the purposes of approximating $ A(\epsilon) $.

The hard ellipse problem consists of two ellipses that are tangent at uniformly random points along their perimeter. We define the polar angle with respect to the minor axis of the ellipses as $ \phi $ and $ \phi' $. By aligning the tangent line slope along the $ y $-axis of the frame of reference, we can calculate the angle of the orientation vector with respect to the horizontal using trigonometry: $ \theta(\phi;\epsilon) = \tan^{-1}(\sqrt{1 - \epsilon^2} \cot \phi) $. Subsequently, we use the difference in orientation angles to find the energy $ \hat u(\epsilon) $ for the two-cell interaction as a function of the eccentricity $ \epsilon $.
\begin{align}
f(\phi,\phi';\epsilon) &= \cos(2\theta(\phi;\epsilon)-2\theta(\phi';\epsilon))\\
\hat u(\epsilon) &= \langle f(\phi,\phi';\epsilon)\rangle\\
&= \frac{1}{\pi^2}\int_0^\pi \int_0^\pi f(\phi,\phi';\epsilon) d\phi d \phi' \\
&= \left(\frac{\sqrt{1 - \epsilon^2} - 1}{\sqrt{1 - \epsilon^2} + 1}\right)^2
\end{align}
To find how this relates to $ A $, we compute the mean energy again, but this time using the Boltzmann distribution from the NCAT Hamiltonian:

\begin{align}
Z(A) &= \int_0^\pi \int_0^\pi \exp(A\cos2(\theta - \theta')) d\theta d\theta' = \pi^2 I_0(A)\\
u(A) &= \frac{\int_0^\pi \int_0^\pi \exp(A\cos 2(\theta - \theta'))\cos 2(\theta - \theta') d\theta d\theta'}{Z(A)}\\
&= \frac{I_1(A)}{I_0(A)}
\end{align}

Setting $ u(A) = \hat u(\epsilon) $ leads to a first order approximation for $ A $ in terms of $ u(\epsilon) $ based solely on the shape of the interacting cells. Since both functions are monotonically increasing, we get $ A \approx (u^{-1} \circ \hat u)(\epsilon) $ and this numerical problem has been solved using piecewise Taylor approximation methods and continued fraction methods. We use the expression from \cite{hill1981evaluation}, evaluated at three separate ranges of $ \hat u(\epsilon) $:
\begin{align}
\alpha(\epsilon) &= \frac{2}{1 - \hat u (\epsilon)}\\
\beta(\epsilon) &= \frac{32}{\alpha(\epsilon) - 131.5 + 120 \hat u (\epsilon)}\\
\gamma(\epsilon) &= 2001 + 4317 \hat u(\epsilon) - 2326 \hat u(\epsilon)^2\\
A &= (u^{-1} \circ \hat u)(\epsilon) \\&\approx 
\begin{cases}
\frac{2\hat u - \hat u^3 - \frac{\hat u^5}{6} - \frac{\hat u^7}{24}+\frac{\hat u^9}{360}+\frac{53\hat u^{11}}{2160}}{(1 - \hat u^2)^{-1}} & \hat u(\epsilon) < 0.85 \\
\frac{1}{4}\left(\alpha + 1 + \frac{3}{\alpha - 5 - \frac{12}{\alpha - 10 - \gamma}}\right) & \hat u(\epsilon) >0.95 \\
\frac{1}{4}\left(\alpha + 1 + \frac{3}{\alpha - 5 - \frac{12}{\alpha - 10 - \beta}}\right) & \mathrm{otherwise}
\end{cases} 
\end{align}
Note that the range of the theoretical \textit{average} energy functions $ u $ and $ \hat u $ is in $ [0,1] $. This approximation works up to 3 decimal places. Finally, we can get an exact analytical expression for $ \epsilon $ in terms of $ A $: $ \epsilon(A) = (\hat u^{-1} \circ u)(A) = 2 \frac{\sqrt{\sqrt{u}(1 + u)-2u}}{1-u}$.

\bibliography{sm}